\documentclass[review,12pt,authoryear]{elsarticle}

\usepackage{url}

\usepackage{amsmath}
\usepackage{amsthm}
\usepackage{amssymb}
\usepackage{bbm}
\usepackage{verbatim}
\usepackage{graphicx}
\usepackage{subfigure}
\usepackage{afterpage}
\usepackage{etoolbox}
\usepackage{float}
\usepackage{rotating}
\usepackage[inline]{enumitem}

\makeatletter
\renewcommand{\p@subfigure}{\thefigure}
\makeatother

\newtheorem{definition}{Definition}[section]
\newtheorem{theorem}{Theorem}[section]

\newtheorem{lemma}[theorem]{Lemma}

\newcommand{\norm}[1]{\left \| #1 \right \|}
\newcommand{\norminline}[1]{\| #1 \|}

\newcommand{\Rn}[1]{{\mathbbm{R}^{#1}}}

\newcommand{\set}{\mathcal}
\newcommand{\vect}{\mathbf}

\newcommand{\map}{\mathbf}
\newcommand{\submat}[3]{{#1}_{({#2},{#3})}}
\newcommand{\rows}[2]{\submat{{#1}}{{#2}}{:}}

\newcommand{\ventry}[2]{{\vect{#1}}_{(#2)}}
\newcommand{\mentry}[3]{{#1}_{(#2,#3)}}

\newcommand{\timel}[2]{{#1}^{(#2)}}

\usepackage{upgreek}
\usepackage{isomath}
\usepackage[algoruled,vlined]{algorithm2e}
\usepackage{algorithmic}

\usepackage{amssymb}

\usepackage{lineno}
\usepackage{verbatim}

\journal{New Astronomy}

\begin{document}

\begin{frontmatter}



\title{Automated identification of transiting exoplanet candidates in NASA Transiting Exoplanets Survey Satellite (TESS) data with machine learning methods}

\author{L. Ofman$^{a,b}$\corref{cor}}
\address[mymainaddress]{Dept. of Physics, Catholic University of America, Washington, DC 20064}
\address[mysecondaryaddress]{NASA GSFC, Code 671, Greenbelt, Maryland, 20771, USA.}
\cortext[cor]{Corresponding author}
\author{Amir Averbuch$^{c,d}$}
\address[mysecondaryaddress]{School of Computer Science, Tel Aviv University, Tel Aviv, Israel}
\author{Adi Shliselberg$^{d}$}
\author{Idan Benaun$^{d}$}
\author{David Segev$^{d}$}
\author{Aron Rissman$^{d}$}
\address[mysecondaryaddress]{ThetaRay, 8 Hanagar Street, Hod HaSharon, Israel}


\begin{abstract}
A novel artificial intelligence (AI) technique that uses machine learning (ML) methodologies combines several algorithms, which were developed by {\em ThetaRay, Inc.}, is applied to NASA's Transiting Exoplanets Survey Satellite (TESS) dataset to identify exoplanetary candidates. The AI/ML {\em ThetaRay} system is trained initially with Kepler exoplanetary data and validated with confirmed exoplanets before its application to TESS data. Existing and new features of the data, based on various observational parameters, are constructed and used in the AI/ML analysis by employing semi-supervised and unsupervised machine learning techniques. By the application of {\em ThetaRay} system  to 10,803 light curves of threshold crossing events (TCEs) produced by the TESS mission, obtained from the Mikulski Archive for Space Telescopes, {the algorithm yields about 50 targets for further analysis, and} we uncover { three} new exoplanetary candidates { by further manual vetting}. This study demonstrates for the first time the successful application of the particular combined multiple AI/ML-based methodologies to a large astrophysical dataset for rapid automated classification of { TCEs}.
\end{abstract}




\end{frontmatter}


\section{Introduction} \label{intro:sec}

The Transiting Exoplanet Survey Satellite (TESS) \citep{Ric14} was launched by NASA on April 18, 2018 with the primary objective of all-sky surveying more than 200,000 near-Earth stars in search of transiting exoplanets using high-precision photometry, producing light curves with a 2-minute cadence. The threshold crossing events (TCEs) list for TESS were produced by the Science Processing Operations Center (SPOC) pipeline at NASA Ames from their search of two-minute light curves. The TESS Objects of Interests (TOI) have been released periodically and archived at the Mikulski Archive for Space Telescopes (MAST, \url{https://archive.stsci.edu/}). The TOI includes planetary candidates, as well as potential planetary candidates and other astrophysical targets, including false positives, comprising the database used for searching for confirmed exoplanets. As of April 20, 2021 TESS has released 2645 TOIs with 122 confirmed planets, 757 TOIs with planetary radius $r_p<R_{Earth}$ and 745 false positives (see, \url{https://tess.mit.edu/publications/}).

Previously, Kepler Space Telescope launched by NASA in 2009 was designed to determine the occurrence frequency of Earth-sized planets. Towards this objective, Kepler observed about 200,000 stars with high photometric precision discovering thousands of transiting exoplanets and exoplanetary candidates \citep{Bor10,Jen10a,Koc10,Chr12}. During the prime missions (2009 May 2 -2013 May 11) Kepler was pointing at a single field of view of about 115 square degrees in the constellations of Cygnus and Lyra. The many periodic signals detected by Kepler were processed using the Kepler Science Processing Pipeline \citep{Jen10b}. They were assembled into a database of TCEs. Direct human input was required to remove false positives and instrumental effects from this database. However, the resulting TCEs database contains data produced by many possible sources, such as eclipsing binaries, background eclipsing binaries and many other possible false alarm sources, in addition to small fraction of exoplanetary candidates (EPCs), and still require considerable analysis for confirmed identification of exoplanets.

Recently, \citet{SV18} identified transiting exoplanets in Kepler satellite data using Deep Learning (DL) algorithm based on training of convolutional neural networks using the Google-Vizier system \citep{Gol17}. \citet{SV18} trained the neural networks to classify whether a given light curve signal is a signature of a transiting exoplanet with low false positive rate. By using their algorithm, they  identify multi-planet resonant chains around Kepler-80. Later, the extended Kepler K2 mission, which starting in Nov. 2013, was designed to use the remaining Kepler capabilities after the completion of the prime mission including the technical failures of the reaction wheels. During this observation phase, the photometric accuracy was reduced, and the pointing varied in different regions of the sky. Nevertheless,  \citet{Dat19} used a similar automated technique based on \citet{SV18} study that is applied to  mission data K2 while identifying two previously unknown exoplanets. Multiple ML algorithms were used to validate exoplanets in Kepler data in other recent studies \citep[e.g.,][]{Armstrong20}. 

Automated classification methods for transiting exoplanets from TESS data have been developed using machine learning (ML) techniques in several studies \citep[e.g.,][]{Zuc18,Yu19,Osb20} that demonstrate the usefulness and feasibility of this approach with various degrees of improved classification performance. In this paper, we describe an application of novel algorithms, which combine several ML approaches and low rank matrix decomposition, including  algorithms that identify anomalies in high dimensional big data by using augmentation approach.  These methods, utilized  semi-supervised and unsupervised learning was developed by {\em ThetaRay, Inc.} (\url{https://thetaray.com/}) for uncovering financial crimes, cyber and Internet of Things (IoT) security, was applied for transiting EPCs search, reported in this study. By using Kepler data with confirmed exoplanets as part of the algorithm training phase and validation, the {\em ThetaRay} platform was applied to TESS data yielding { three} new EPCs out of nearly 11000 TCEs, demonstrating the feasibility and utility of this new platform with unsupervised ML methods.

 The paper is organized as follows: Section~\ref{ML:sec} discusses the ML methods and the details of {\em ThetaRay} algorithms. Kepler Satellite data ML training are described in Section~\ref{Kepler:sec}.  Section~\ref{TESS:sec} is devoted to the description TESS Satellite Data Analysis and the results of application of {\em ThetaRay} unsupervised ML algorithm to TESS data. Section~\ref{DC:sec} is devoted to the discussion and conclusions.

\section{Machine Learning Methods}
\label{ML:sec}
\subsection{The {\em ThetaRay} Algorithm}
In the present study we utilize  {\em ThetaRay} AI-based Fintech algorithms, commercially developed for anomaly detection (financial crimes) in financial institutions, cyber security and IoT for smooth operations of critical infrastructure installations. Since transiting exoplanets light curves are rare and only appear in small number of all observed  Kepler or TESS stellar light curves, they are classified as ‘anomalies’ in our analysis when {\em ThetaRay} system utilizes  the strengths of its algorithms to identify transiting EPCs in the large number of TCEs.  To identify these ‘anomalies’, or exoplanet light-curves, {\em ThetaRay}’s algorithms generates a data-driven ‘normal’ profile of the data ingested, and simultaneously identifies anomalies also called abnormal events, providing forensics that categorizes each event based on its features. This is done autonomously by the algorithm without the need to have rules or signatures. {\em ThetaRay}’s algorithmic engine utilizes techniques drawn from a wide variety of mathematical disciplines, such as harmonic analysis, diffusion geometry and stochastic processing, low rank matrix decomposition, randomized algorithms in general and randomized linear algebra in particular, geometric measure theory, manifold learning, neural networks/deep learning, and compact representation by dictionaries. One approach models the data as a diffusion process using Brownian motion of a random walk process to geometrize the data. There is no need for any semantic understanding of the processed data, nor are there any predefined rules, heuristics or weights in the system. The diffused collected dataset is then converted into a Markov matrix through a normalized graph-Laplacian and modeled as a stochastic process that is applied in many dimensions (could reach thousands).

The classification of light curves as exoplanetary candidates in this paper is achieved by using the analytic platform of  {\em ThetaRay} that is described below. This platform processes high dimensional big data to identify anomalous behavior in comparison to a normal profile. This anomaly detection tool is used in the present application for classification of EPCs in TESS TCE database. The normal profile is a training data driven and its generation is explained below. In the present study we used Kepler TCE data as a training dataset as described in section~\ref{Kepler:sec}.

Below, we describe some of the algorithms that were utilized in the study of identifying anomalies in a big data using augmentation, semi-supervised and unsupervised type algorithms. The same core algorithms for anomaly identification are capable of identifying anomalies in cyber (malware), industrial malfunction  (IoT) and financial (crimes) data.  The algorithms were applied for the first time to astrophysical data in this study. These algorithms are part of {\em ThetaRay} (\url{www.thetaray.com}) core technology portfolio to fight financial crimes \citep{shabat2018uncovering}. The algorithms are housed in {\em ThetaRay} Computational Platform that enables efficient data manipulation and processing. The reported results were obtained by executing these algorithms on {\em ThetaRay} platform described below.

\subsection{Semi-supervised processing via augmentation: Introduction}
For background and context, we describe briefly the {\em ThetaRay} system  current commercial applications that now have been expanded and applied to astrophysics dataset. The {\em ThetaRay} is designed to provide     fast and accurate analytic solutions for identifying emerging risk/crime (classified as anomalies) in financial data, discovering new opportunities, and exposing blind spots within these large, complex high dimensional data sets.  These AI-based algorithms radically reducing false positives and are uniquely able to uncover ``unknown unknowns” (these are threats that one is not aware of, and do not even know that one is not aware of them). {\em ThetaRay} provides constructive solutions to anomaly detections challenges via  its analytic platform designed for a big data,  uncover previously unknown risks, and do so with industry low false positive rates and in real time enabling fast forensic.

In this project, we assume that some labels of Kepler TCE data, which is a related dataset to TESS TCEs, are given but are not given for the TESS data. An augmented algorithm, which is considered as a learning method,  generates a new data frame based on the provided labels. Then, the new data frame serves as an input to unsupervised algorithms. In this project, we apply 3 unsupervised algorithms to the augmented data: Geometric-based denoted by NY (see section \ref{NY1}), algebraic-based denoted by LU (see section \ref{LU-dictionary}), a hybrid of LU and NY and Neural network denoted by  AE.

The augmentation method is based on Neural Network.
By using a Neural Network-based method, the default network (that can be user-adjusted) consists of one input layer (the analysis data frame), three hidden layers and one output layer. All the layers are connected through ``weights” that are automatically tuned during the learning (optimization) process until the network output layer values are close to the values of the provided labels. After optimization, the third hidden layer becomes the new data frame as well as the input to the unsupervised algorithms that are outlined in section  \ref{list-of-algorithms} and  some of them are described in detail in section \ref{unsupervised-detailed-description}.

{\em ThetaRay}'s platform covers detection and monitoring of several verticals with current emphasis on financial crimes by suppling an end-to-end solution. {\em ThetaRay} provides  an un- and semi-supervised  real-time diagnostic, AI based financial crimes detection platform that are based on anomaly detection algorithms of ``unknown unknowns".

Rule-based technology, which is very popular among anomaly detection tools,  is intended for what is known and when you know what to look for. {\em ThetaRay}’s detection is achieved by un-  and semi-supervised  with automatic methods that are not based on  rules, patterns, signatures, heuristics, data semantics of the features or any prior domain expertise and provide high detection rate and very low false positives.  {\em ThetaRay}'s methodologies within its Analytics Platform are based on unbiased detection through a series of randomized advanced AI-based algorithms that can process any number of data features and can be explained, justified and anomalies can be traced back to identify features that triggered the anomalies therefore it is not classified as a black box. Thus, the platform enables past tracking of events and features that trigger the occurrence of anomalies.
{\em ThetaRay}'s system operates under the assumption that is not know what to look for or what to ask. This allows their technology to potentially, detect every type of anomaly before the rules are discovered automatically. For efficient processing of the algorithms the system uses off-the-shelf hardware components. Inherent parallelism in the algorithms is  implemented with GPU utilization. The platform contains advanced and interactive visualization of the input and output phases of the data analysis. The detection approach is data driven thus;  no preexisting models are assumed to exist. This makes this approach  universal and generic and thus opens the way for different applications without the introduction of bias, limitations, and unfounded preconceptions into the processing, a property well suited for large astrophysical datasets. Mathematical and physical justification for most of the available algorithms in the system are given below.

The input training data can be enriched by a given limited set of labels. This increases the detection rate and reduces the false alarm rates. This is part of semi-supervised algorithms. Semi- and un-supervised algorithms are used. Currently, the platform contains eight different unsupervised algorithms for the data without labels and three different semi-supervised algorithms for the data with partial labels within the detection engine.  The results are fused to produce one solution. {\em ThetaRay} combines the strengths of unsupervised and semi-supervised techniques to identify anomalies in the data. Unsupervised learning assumes that there are no labels to the various data components. Semi-supervised learning frameworks have made significant progress in training machine learning with limited labeled data in image domain. Augmented unsupervised learning can be used side-by-side with semi-supervised learning. The augmented algorithms generate a new data frame based on the analysis data frame and the provided labels. The new data frame generated is then the new input for all the unsupervised algorithms selected. Labels are categorized as binaries, with the minority of the labels (known anomalies) marked as ``1” and the remainder, which are the majority of unknown cases, assigned ``0”.

Augmented process enables covering both the known and the unknown with a relative balance between them. The {\em ThetaRay} system allows for configuration of the underlying input features, algorithms and detection logic at each application.  Technically it is a neural network-based process which generates a new data frame based on the input data frame and binary labels provided by the application (in the present case, stellar light-curve data).

\subsection{Unsupervised algorithms: General description}
\label{list-of-algorithms}
\begin{description}
\item[NY:]
This algorithm (see, Figure~\ref{NYflow:fig}) is based on diffusion maps (DM) methodology \citep{lafon:DM} and it is primarily a non-linear dimension reduction process. The anomaly identification procedure takes place inside the lower dimensional space (manifold) that is determined automatically during the training phase. An out-of-sample extension procedure \citep{coifman2006geometric} is applied to the identification phase for each multidimensional data point, which did not participate in the training phase, to determine whether it belongs to the manifold (low dimensional space - classified as normal) or deviates from it (classified as anomalous).
\begin{figure}[ht]
 \centering
\includegraphics[width=\textwidth]{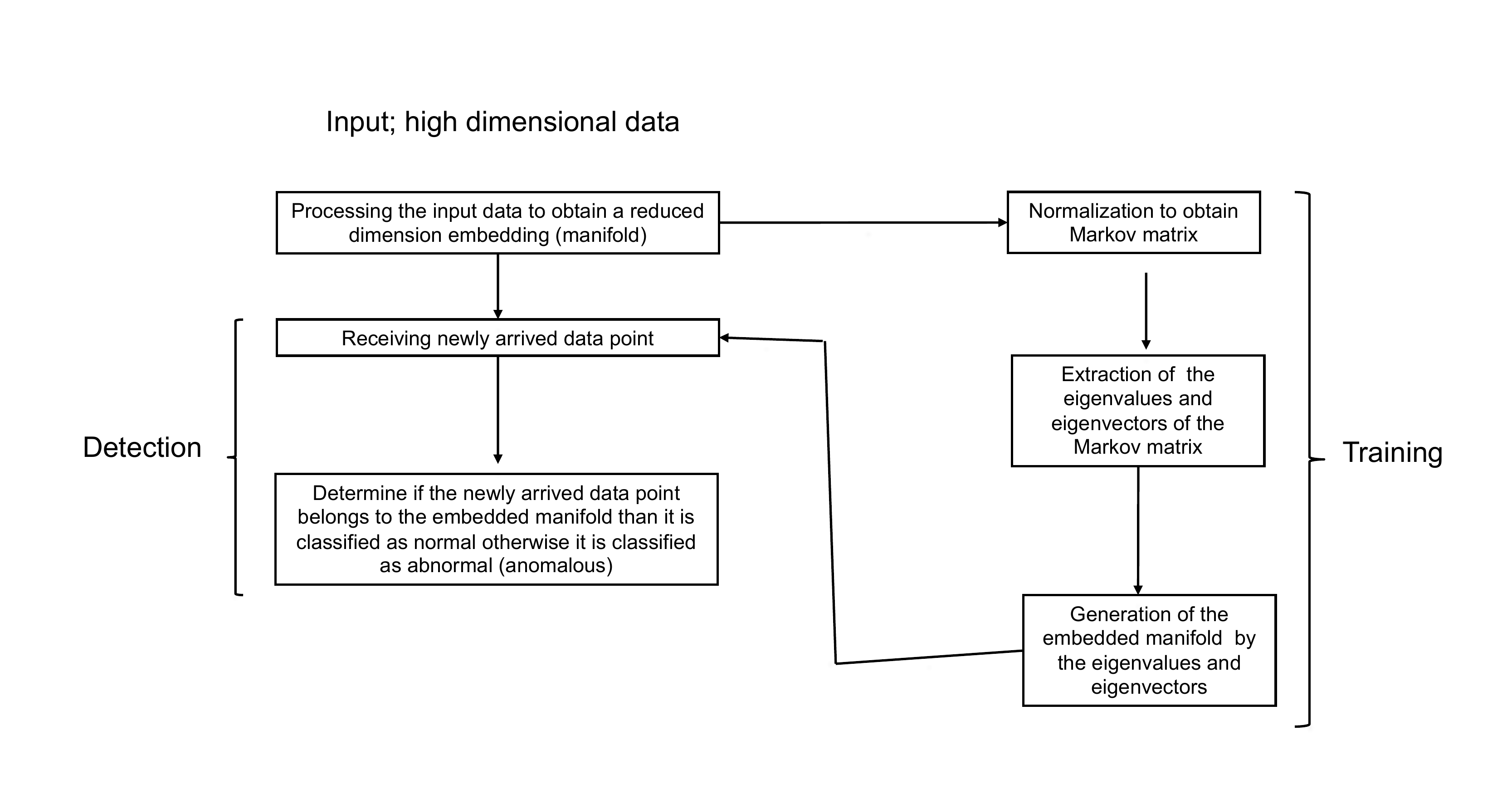}
 \vspace{-1.5cm}\caption{NY algorithm: flow chart.}
    \label{NYflow:fig}
\end{figure}

The NY algorithm, which is based on DM, geometrizes the input training data. DM analyzes the ambient space (training data) and determines automatically where the data actually resides in the embedded space.  We can visualize the input training data (ambient space) as a matrix of size $m \times n$ where $m$ is the number of multidimensional data points (number of rows in the matrix) and each row is of dimension $n$ - the number of columns in the matrix. The input data is assumed to be sampled from a low dimensional manifold (embedded space) that captures the dependencies between the observable parameters. DM reduces in a non-linear way the dimension of the ambient space which is the training data. The dimensionality reduction by DM  is based on local affinities between multidimensional data points and on non-linear embedding of the ambient space into a lower dimensional space, described as a manifold, by using a low rank matrix decomposition. The non-parametric nature of this analysis uncovers the important underlying factors of the input data and reveals the intrinsic geometry of the data represented by the embedded manifold. This manifold describes geometrically what we classify as the normal profile of the ambient data. Newly arrived multidimensional data points, which did not participate in the training procedure, are embedded into the lower dimensional space by the application of an out-of-sample extension algorithm. If the embedded multidimensional data point falls into the manifold, it is classified as normal otherwise it is classified as abnormal (anomalous).
See section \ref{NY1} for more details.
\item[LU:]
Based on a randomized low-rank matrix decomposition \citep{shabat2016randomized}. This algorithm builds a dictionary from the training data. Then, each newly arrived multidimensional data point that is not well described (not spanned well) by the dictionary is classified as an anomalous data point.

The randomized LU (RLU) algorithm is an algebraic approach applied to input matrix $A$ of size $m\times n$
with an intrinsic dimension $k$ smaller than $n$. $k$ can be computed automatically
or given. RLU is a low rank matrix decomposition which enables the identification of anomalies
using a dictionary constructed from the training data. RLU forms a low rank
matrix approximation of $A$ such that $PAQ \approx LU$ where $P$ and $Q$ are orthogonal permutation matrices,
and $L$ and $U$ are the lower and upper triangular matrices, respectively. A dictionary
is then constructed according to $D = P^{T}L$ ($T$ is the transpose of a matrix). Thus, $D$ is a linear
combination of the input matrix and a representation of the normal data. It is also used in
the identification step to classify newly arrived multidimensional data points that did not
participate in the training phase. Thus, a new incoming a multidimensional data point $x$, which satisfies $\|DD^{\dagger}x-x\|<\epsilon$, is classified as normal; otherwise, it is classified as anomalous. Here, $D^{\dagger}$ is the
pseudo inverse of $D$ and $\epsilon$ is a quantity defined in the training phase. When applied to a matrix $A$ of size $m \times n$, the RLU decomposition reduces the number $m$ of
multidimensional data points, resulting in a reduced-measurements matrix 
of size $k\times n$
where $k<n<m$. Although the algorithm is a randomized, it has been proven in \citet{shabat2016randomized} that the probability that the RLU approximation  will generate  a big error tends to be very small.
See section \ref{LU-dictionary} for more details.

\item[DK:]
The DK Algorithm relies on successive applications of LU and NY. Assume the size of a given training matrix is $m$ data points (rows) by $n$ features (columns). RLU (described in section \ref{LU-dictionary}) is applied to $n$. The size of $n$ is reduced substantially through the application of random projection \citep{johnson1984extensions}.
Then, NY (described in section \ref{NY1}) is applied to $n$ (dimension) and the matrix is embedded into a lower dimensional space and anomaly identification procedure NY is called in this embedded space.

\item[AE:]
This is a variational autoencoder (AE) algorithm.
AE is machine learning tool designed to generate complex models of data after careful distribution modeling of example data. In neural net language, AE consists of an encoder component and a decoder component. We assume that the input data set is generated from an underlying unobserved (latent) representation. Given an input data set, the encoder part of an AE approximates the distribution of the latent variables. Finally, the algorithm sets the distribution parameters of the latent layers in a manner that maximizes the likelihood of generating or reconstructing the input data in the decoder section.
As soon as the distribution of the latent variables is approximated, we can sample from this distribution to generate an approximate representation of the input data. Since normality consists of and is defined by most of the data points, those will be well-approximated by the AE, while anomalies will be poorly modeled. Therefore, by comparing the original sample with the reconstructed (generated) data, we can calculate a similarity score that enables us to detect anomalies.
The goal is to use the AE as a denoising autoencoder. It allows us to encode our sample into the latent space and then reconstruct it. By comparing the original sample to the reconstruction, we are able to calculate a score that enables us to classify a data point as  anomalous  data point.
Since we plan to use the AE for anomaly detection, we have to  calculate the scores for the input and output.


\end{description}

\subsection {Unsupervised algorithms: Mathematical description}
\label{unsupervised-detailed-description}

\subsubsection{Diffusion geometry: Background}
\label{NY1}

DM 
 are a kernel-based
method for manifold learning that can reveal the intrinsic structures
in data and embed them in a low dimensional space.  The DM-based approach computes the
diffusion geometry. A spectral embedding of the data points
provides coordinates that are used to interpolate and
approximate the pointwise diffusion map embedding of data.

Manifold learning approaches are often used for modeling
and uncovering intrinsic low dimensional structure in high
dimensional data. DM is a method that captures data manifolds
with random walks that propagate through non-linear pathways
in the data. Transition probabilities of a Markovian diffusion
process (explained later how to compute them) define an intrinsic diffusion distance metric that is
amenable to a low dimensional embedding. By arranging
transition probabilities in a row-stochastic diffusion operator,
and taking its leading eigenvalues and eigenvectors, one can
derive a small set of coordinates where diffusion distances are
approximated as Euclidean distances and intrinsic manifold
structures are revealed.

In more details, the NY  algorithm uncovers the internal geometry of the input training data denoted as $A$. The use of geometric considerations speeds up significantly the anomaly detection computational time.
Next is a theory that supports this approach: The goal is to detect anomalies in $A$ and in newly arrived $n$-dimensional data points that did not participate in the training data $A$.  During the training procedure,  size of $n$,  which is also called the dimension of $A$, is automatically reduced. The procedure is called dimensionality reduction. Dimensionality reduction as explained later, is achieved without damaging the quality and the coherency of the data in $A$. More than that, there is no loss of data as explained later. Dimensionality reduction is just a different representation of the training data that automatically without any human intervention reduced the dimension according to the data and uncovers the real dimension where the training data actually resides.

In general, anomaly detection  is based on the notion of similarities (or affinities) between the $m$ high dimensional data points (these are the rows in the matrix $A$). How we detect anomalies in this big data efficiently without introducing bias and without damaging the data? Dimensionality reduction of $n$ is needed. How to achieve this reduction?  The following provides the rationale why geometrization of the training data $A$ and tracking the movement of newly arrived data points identify a low dimensional manifold for learning. It is founded mathematically through the preservation of the quality and the integrity (completeness) of the data in $A$.

The assumption is that the processed data is imbalance: High densities of $n$-dimensional samples (rows in the matrix $A$)  represent normal data otherwise the data is classified as anomalous (abnormal) since the majority of the data is normal and thus it is classified as having high density.

Theory: How to find the low dimensional space (manifold)? It is proved that if $A$ is sampled from a low intrinsic dimensional manifold, then, as $n$ (dimension) tends to infinity, the defined random walk, which travels between all the data samples, converges to a diffusion process over the manifold. This is the key to the processing of $A$ as diffusion process that guarantees efficient scan of the data through randomization without introduction of bias. It provides three complementary approaches for dimensionality reduction – diffusion distances between $n$-dimensional samples, randomization and manifold learning -  emerge from this observation (theorem): 1.  $A$ kernel matrix $B$ of size $m\times m$ (huge)  is constructed from distances among all the $n$-dimensional samples (rows). The distances are diffusion distances. 2.  Random walk is applied to the entries in $B$.  This random walk guarantees that there is no bias between the utilization of the distances in $B$. 3. Diffusion Maps  (DM) links between the matrix $B$ and  a lower dimensional space (manifold)  via diffusion processing. The dimension of the embedded manifold represents the reduction of  $n$.

Geometrization of the training data - outline description of the approach: The { NY} algorithm is based on a geometric uncovering of a low dimensional  manifold in the ambient space (the original space represented by $A$)  by the application of DM to ambient space represented by $A$. The input data is assumed to be sampled from a low intrinsic dimensional manifold that captures the dependencies between the observable parameters ($n$-dimensional features). DM reduces the dimension $n$ of the training data. It is based on local affinities between multidimensional data points and on non-linear embedding of the ambient space into a lower dimensional space, described as a manifold, by using a low rank matrix decomposition. The non-parametric nature of this analysis uncovers the important underlying factors of the input data and reveals the intrinsic geometry of the data represented by the embedded manifold. This manifold describes geometrically what we classify as the normal profile in the ambient data. Newly arrived n-dimensional data points, which did not participate in the training procedure, are embedded into the lower dimensional space by the application of an out-of-sample extension algorithm. If the embedded n-dimensional data point falls into the manifold where most of the normal data reside, it is classified as normal; otherwise, it is classified as abnormal (anomalous).
The exchange of  data between the ambient space and the manifold, where the detection takes place, does not degrade the coherency and the completeness of the data and preserves the geometrical relations (affinities) between the two spaces – ambient and embedded (manifold).

\subsubsection{Diffusion geometry: outline}

 Let $\set X=\{x_1,\ldots,x_n\}$ be a dataset
and let $\map k:\set X\times \set X\to\Rn{}$ be a symmetric
point-wise positive kernel that defines a connected, undirected and
weighted graph over $\set X$. Then, a random walk over $\set X$ is
defined by the $n\times n$ row-stochastic transition probabilities
matrix
$P = D^{-1}K$,
where $K$ is an $n\times n$ matrix whose entries are
$\mentry{K}{i}{j}:=\map k(x_i,x_j),~i,j=1,\ldots,n,$ and $D$ is the
$n\times n$ diagonal degrees  matrix whose $i$-th element is $
\ventry{d}{i}:=\sum_{j=1}^n \map k(x_i,x_j),~i=1,\ldots,n. $ The
vector $\vect d\in\Rn{n}$ is referred to as the \emph{degrees
vector} of the graph defined by $\map k$.

The associated time-homogeneous random walk $X(t)$, is defined via the conditional probabilities on its state-space $\set X$: assuming that the process starts at time $t=0$, then for any time point $t\in\mathbb N$
$\mathbbm{P}(X(t) = x_j | X(0)=x_i) = \mentry{P^t}{i}{j}$,
where $\mentry{P^t}{i}{j}$ is the $(i,j)-$th entry of the $t$-th power of the matrix $P$. As long as the process is aperiodic, it has a unique stationary distribution $ {\vect {\hat d}}\in\Rn{n}$ which is the steady state of the process, i.e. ${\ventry{\hat d}{j}} = \lim_{t\to\infty}\mentry{P^t}{i}{j}$, regardless the initial state $X(0)$. This steady state is the probability distribution resulted from $\ell_1$ normalization of the degrees vector $\vect d$, i.e.,
\begin{equation}
\label{eq:d_hat}
\vect{\hat d} = \frac{\vect d}{\norminline{\vect d}_{1}}\in\Rn{n},
\end{equation}
where $\norminline{\vect d}_{1}:=\sum_{i=1}^n\ventry{d}{i}$. The \emph{diffusion distances} at time $t$ are defined by the metric $\timel{\map D}{t}:\set X\times \set X\to\Rn{}$,
\begin{eqnarray}
\label{eq:DM}
\timel{\map D}{t}(x_i,x_j)&:= &\norm{\rows{P^t}{i}-\rows{P^t}{j}}_{\ell^2(\vect{\hat d}^{-1})}\nonumber\\
& = &\sqrt{\sum_{k=1}^n(\mentry{P^t}{i}{k}-\mentry{P^t}{j}{k})^2/\ventry{\hat d}{k}},~i,j=1,\ldots,n.
\end{eqnarray}
By definition, $\rows{P^t}{i}$, the $i$-th row of $P^t$, is the
probability distribution over $\set X$ after $t$ time steps given
that the initial state is $X(0)=x_i$. Therefore, the diffusion
distance $\timel{\map D}{t}(x_i,x_j)$ from Eq.~\ref{eq:DM} measures
the difference between two propagations along $t$ time steps: the
first is originated in $x_i$ and the second in $x_j$. Weighing the
metric by the inverse of the steady state results in ascribing high
weight for similar probabilities on rare states and vice versa.
Thus, a family of diffusion geometries is defined by
Eq.~\ref{eq:DM}, each corresponds to a single time step $t$.

Due to the above interpretation, the diffusion distances are
naturally utilized  for multiscale clustering since they uncover the
connectivity properties of the graph across time. In~\citet{berard,
lafon:DM} it has been proven that under some conditions, if
$\set{X}$ is sampled from a low intrinsic dimensional manifold, then,
as $n$ tends to infinity, the defined random walk converges to a
diffusion process over the manifold.

\subsection{Randomized LU decomposition: An algorithm for dictionary construction}
\label{LU-dictionary}

A dictionary construction algorithm is presented. It is based on a low-rank matrix
factorization being achieved by the application of the randomized LU
decomposition \citep{shabat2016randomized} to a training data. This method is fast, scalable, parallelizable,
consumes low memory, outperforms  SVD in these categories and works
also extremely well on large sparse matrices. In contrast to
existing methods, the randomized LU decomposition constructs an
under-complete dictionary, which simplifies both the construction
and the classification processes of newly arrived multidimensional data points. The
dictionary construction is generic and general  that fits
different applications.

The randomized LU algorithm, which is applied to a
given training data matrix $A\in\mathbb{R}^{m\times n}$ of $m$ multidimensional data points  and $n$ features, decomposes $A$ into
two matrices $L$ and $U$. The size of $L$ is determined by the decaying spectrum of the singular
values of the matrix $A$, and bounded by $\min\{n,m\}$. Both
$L$ and $U$ are linearly independent.

\begin{figure}
    \centering
       \includegraphics[width=\textwidth]{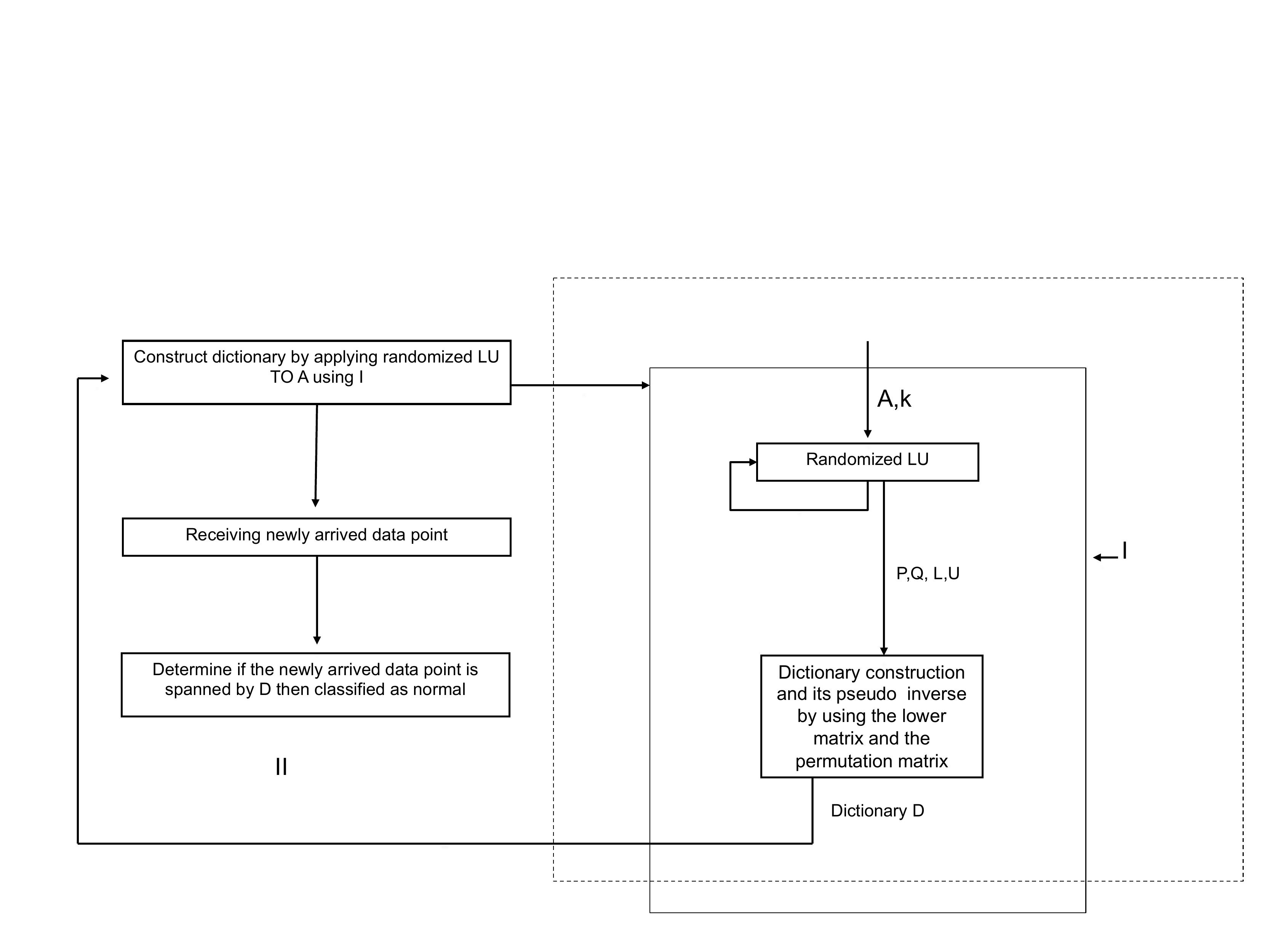}
 \caption{II calls the construction of  a  dictionary $D$ via randomized LU decomposition as described in I. The LU algorithm is built from the following steps: The inputs to the algorithm are  a matrix $A$ and its rank $k$ (see I). They are submitted to Randomized LU that generates the following outputs: Permutation matrices $P$ and $Q$ and lower and upper triangle matrices $L$ and $U$, respectively. Then, a newly  arrived data point, that did not participate in the training, is either spanned by $D$ therefore classified as normal otherwise it is classified as abnormal (anomalous). }
    \label{LUflow}
\end{figure}

The randomized LU decomposition algorithm (see, Figure~\ref{LUflow}) computes the rank $k$ LU approximation
of a full matrix (Algorithm \ref{fileRec:alg:randomized_lu}).
The main building blocks of the algorithm are random projections and Rank Revealing LU (RRLU)
\citep{pan2000existence} to obtain a stable low-rank approximation
for an input matrix $A$ that is classified as a training data. In  Figure~\ref{LUflow} `II' describes the generation of a dictionaries by calling  item I that describes the flow of the randomized  LU decomposition. The end of the execution of `I' means  that the training is completed. The dictionaries are the input of `II' that performs the identification. Newly  arrived data point that did not   participate in the training is either span (classified as normal) or not spanned by the dictionary (classified as anomalous).

The RRLU algorithm, used in Algorithm~\ref{fileRec:alg:randomized_lu}, reveals the
connection between LU decomposition of a matrix and its singular values.
 Similar algorithms exist for
rank revealing QR decompositions (see, for example \citet{gu1996efficient}).

\begin{theorem}[\citet{pan2000existence}]
\label{fileRec:trm:rrlu-pan}
Let $A$ be an $m\times n$ matrix ($m \gg n$).
Given an integer $1 \le k < n$, then the following factorization
\begin{equation}
\label{fileRec:eq:rrlu_def}
PAQ = \begin{pmatrix} L_{11} & 0 \\ L_{21} & I_{n-k} \end{pmatrix}
\begin{pmatrix} U_{11} & U_{12} \\ 0 & U_{22} \end{pmatrix},
\end{equation}
holds where $L_{11}$ is a lower triangular with ones on the diagonal, $U_{11}$ is an upper triangular, $P$ and $Q$ are orthogonal permutation matrices.
Let $\sigma_1 \ge \sigma_2 \ge ... \ge \sigma_n \ge 0$ be the singular values of $A$, then:
\begin{equation}
\label{fileRec:eq:rrlu_def1}
\sigma_k \ge \sigma_{min}(L_{11}U_{11}) \ge \frac{\sigma_k}{k(n-k)+1},
\end{equation}
and
\begin{equation}
\label{fileRec:eq:rrlu_def2}
\sigma_{k+1} \le \Vert U_{22} \Vert \le (k(n-k)+1)\sigma_{k+1}.
\end{equation}
\end{theorem}

Based on Theorem \ref{fileRec:trm:rrlu-pan}, we have the following definition:
\begin{definition}[RRLU Rank $k$ Approximation denoted RRLU$_k$]
Given a RRLU decomposition (Theorem \ref{fileRec:trm:rrlu-pan}) of a matrix $A$ with an integer $k$ (as in Eq. \ref{fileRec:eq:rrlu_def})
such that $PAQ=LU$, then the RRLU rank $k$ approximation is defined by taking $k$ columns
from $L$ and $k$ rows from $U$ such that
\begin{equation}
\label{fileRec:eq:rrlu_rank_k_approx}
\text{RRLU}_k(PAQ)=\begin{pmatrix} L_{11} \\ L_{21} \end{pmatrix}
\begin{pmatrix} U_{11} U_{12} \end{pmatrix}.
\end{equation}
where $L_{11}, L_{21}, U_{11}, U_{12}, P$ and $Q$ are defined in Theorem \ref{fileRec:trm:rrlu-pan}.
\end{definition}

\begin{lemma}[~\citet{shabat2016randomized} RRLU Approximation Error]
The error of the RRLU$_k$ approximation of $A$ is
\label{fileRec:lem:rrlu_approx_err}
\begin{equation}
\Vert PAQ-\text{RRLU}_k(PAQ) \Vert \le (k(n-k)+1)\sigma_{k+1}.
\end{equation}
\end{lemma}

Algorithm~\ref{fileRec:alg:randomized_lu} describes the flow of the RLU decomposition algorithm.

\begin{algorithm}
\caption{Randomized LU Decomposition}
\label{fileRec:alg:randomized_lu}
\textbf{Input:} Matrix $A$ of size $m \times n$ to decompose; $k$  rank of $A$; $l$ number of columns to use (for example, $l=k+5$).\\
\textbf{Output:} Matrices $P,Q,L,U$ such that $\Vert PAQ-LU \Vert \le \mathcal{O}(\sigma_{k+1}(A))$ where $P$ and $Q$ are orthogonal permutation matrices,
$L$ and $U$ are the lower and upper triangular matrices, respectively, and $\sigma_{k+1}(A)$ is the ${(k+1)}$th singular value of $A$.
\begin{algorithmic}[1]
\STATE Create a matrix $G$ of size $n \times l$ whose entries are
i.i.d. Gaussian random variables with zero mean and unit standard deviation.
\STATE $Y \gets AG$.
\STATE Apply RRLU decomposition (See \citet{pan2000existence}) to $Y$ such that $PYQ_y=L_yU_y$.
\STATE Truncate $L_y$ and $U_y$ by choosing the first $k$ columns and $k$ rows, respectively: $L_y \leftarrow L_y(:,1:k)$ and $U_y \leftarrow U_y(1:k,:)$.
 \STATE $B \gets L_y^{\dagger}PA$. ($L_y^{\dagger}$ is the pseudo inverse of $L_y$).
\STATE Apply LU decomposition to $B$ with column pivoting $BQ=L_bU_b$.
\STATE $L \gets L_y L_b$.
\STATE $U \gets U_b$.
\end{algorithmic}
\end{algorithm}

%


\subsection{Randomized LU Based Classification Algorithm}

Based on Section~\ref{LU-dictionary}, we apply the randomized LU decomposition (Algorithm~\ref{fileRec:alg:randomized_lu}) to matrix $A$, yielding
$P A Q \approx L U$.
 The outputs $P$ and $Q$ are orthogonal permutation matrices.
Theorem~\ref{fileRec:trm:dictionaryerr} shows that $P^T L$ forms (up to a
certain accuracy) a basis to $A$. This is the key property of the
classification algorithm.
\begin{theorem}[~\citet{shabat2016randomized}]

\label{fileRec:trm:dictionaryerr} Given a matrix $A$. Its randomized LU
decomposition is $PAQ \approx LU$. Then, the error of representing
$A$ by $P^T L$ satisfies:
\begin{equation}
\Vert (P^T L)(P^T L)^\dagger A-A \Vert \le \left(2\sqrt{2nl\beta^2\gamma^2+1}+2\sqrt{2nl}\beta\gamma \left( k(n-k)+1 \right)\right)\sigma_{k+1}(A).
\end{equation}
\end{theorem}

Let $x$ be a multidimensional data point  and $D=P^{T}L$ is a dictionary. The
distance between $x$ and the dictionary $D$ is defined by
$dist(x,D) \triangleq || DD^{\dag}x - x||$,
where $D^{\dag}$ is the pseudo-inverse of the matrix $D$. If $dist(x,D)\leq \epsilon$ then $x$ is normal otherwise it is anomalous. \emph{}








The effectiveness of the algorithm and its evaluation are demonstrated om Kepler data in Table \ref{performance-kepler}  at identifying known planet candidates.
The analysis was done in an unsupervised mode using our augmentation methodology. Accuracy is measured by AUC. 

\begin{table}[H]
  \centering
  \begin{tabular}{l l l l l}
     \hline
    True positive & False positive & True negative &  False negative & Total \\
    \hline
     \multicolumn{5}{c}{Training} \\
         2883 & 52 & 14 & 9660 & 12609  \\
         \multicolumn{5}{c}{AUC=0,957, 80\% of the data was used.}\\
         \hline
         \multicolumn{5}{c}{Detect} \\
         622 & 126 & 81 & 2299 & 3128  \\
         \multicolumn{5}{c}{AUC=0,921, 20\% of the data was used.}\\
     \hline
   \end{tabular}
  \caption{The effectiveness of the algorithm and its evaluation on Kepler data of size 15,737. False negative in the detect`  is 2.5\%}
  \label{performance-kepler}
\end{table}

\subsection{Augmentation for unsupervised learning}
The augmented algorithm generates a new data frame when labels are provided. The new data frame then serves as the input to the unsupervised algorithms. The new data frame  is defined as a separate learning method. The provided labels are categorized as binaries where the minority of the labels (known as anomalies) are marked as ``1” and the remainder, which are the majority of unknown cases, is marked as ``0”.
Two methods are available for achieving augmentation: 1. Based on neural network. 2. Based  on geometric component (GC) \cite{bermanis2021geometric} analysis combined with SVM. 
For the neural network-based method, the network consists of one input layer (the  data frame), several hidden layers and one output layer. All the layers are connected through ``weights” that are automatically tuned during the learning (optimization) process, until the network output layer values are close to the values of the labels provided. After optimization, the third hidden layer becomes the new data frame, as well as the input for the unsupervised algorithms described before. 
The GCSVM method first applies the GC algorithm to geometrically down sample the ``0” majority class. This is followed by an auto tuned SVM classifier which segregates between the two classes based on the labels provided. The classification information is then used to generate a new data frame for further unsupervised anomaly detection.

Geometric Component (GC) Analysis provides a dictionary-based framework for geometrically driven data analysis, for both linear and non-linear (diffusion geometries), that includes dimensionality reduction, out of sample extension and anomaly detection. The main algorithm greedily picks multidimensional data points that form linear subspaces in the ambient space that contain as much information as possible from the original data. The GC-based diffusion maps appear to be a direct application of a greedy algorithm to the kernel matrix constructed in diffusion maps. The algorithm greedily selects data points from the data according to their distances from the subspace spanned by the previously selected data points. When the distance of all the remaining data points is smaller than a prespecified threshold, the algorithm stops. The extracted geometry of the data is preserved up to a user-defined distortion rate. In addition, a subset of landmark data points, known as dictionary, is identified by the presented algorithm for dimensionality reduction that is geometric-based. It achieves good results for unsupervised learning tasks. The proposed algorithm is attractive for its simplicity, low computational complexity and tractability.

\subsection{Working on {\em ThetaRay}'s System}
We built in {\em ThetaRay} platform an ``analysis chain”, which is a multi-staged flowchart, that is composed of three main stages: Data Source, Data Frame and Analysis. The data is organized into  data sources and they are uploaded to {\em ThetaRay}'s platform. We created data frames in the system with wrangling method (where, data wrangling is a process of cleaning, structuring and enriching raw data into a desired format with the intent of making it more appropriate and valuable for modeling) and split the data randomly into train and test in {\em ThetaRay} system such that 80\% is allocated for training and 20\% are allocated for testing. The training procedure generates profile and this was fed into different types of  analyses using {\em ThetaRay} Augmented and unsupervised algorithms, to find the best parameters that maximize the Area Under ROC Curve (AUC) in each chain, where ROC is Receiver Operating Characteristic (ROC) curve - a standard evaluation metrics for testing classification model’s performance. After the analysis  and review of these results were completed, the data was processed again after modification and fine tuning of the internal parameters in the system for results improvement. Then, the identification was executed again.

\begin{figure}[H]
    \centering
       \includegraphics[width=\textwidth]{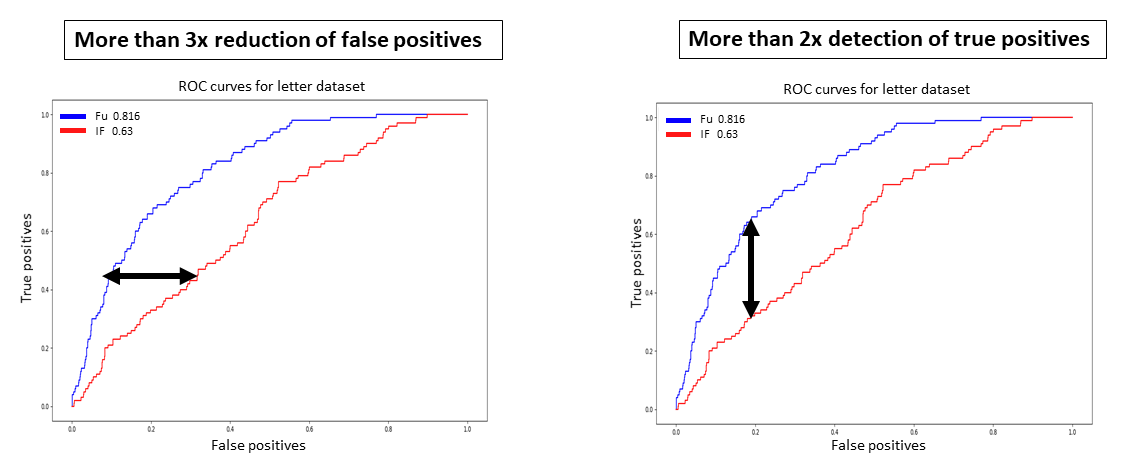}
 \caption{Left: More than 3x reduction of false positives. Right: More than 2x detection of true positives. On both graphs, the AUC of the fusion (Fu) of {\em ThetaRay}’s algorithms is 0.816 while Isolation Forest (IF) AUC is 0.63}
    \label{comparison}
\end{figure}
We compared {\em ThetaRay} algorithms’ performance to a leading open-source unsupervised algorithm. They were applied to 17 datasets. Isolation Forest \citep{liu2008isolation}  was chosen as the leading open-source algorithm. The performance comparison is displayed in Figure~\ref{comparison} where red represents the performance of Isolated Forest (IF) and blue represents the fused performance (Fu) by {\em ThetaRay}’s algorithms.  The comparison was done between our results that were derived from fusion of the participated  algorithms (denoted as Fu) with the Isolated Forest (denoted as IF). The improved performance of the {\em ThetaRay}’s Fu algorithms over the IF algorithm is evident.

\section{Kepler Satellite data ML training}
\label{Kepler:sec}
We have focused on light curves produced by the Kepler space telescope, which collected the light curves of $\sim$200,000 stars in our milky way galaxy for 4 years with continuous 30-min or 1-min sampling. To train the algorithm to identify planets candidates in Kepler light curves, we used a training set of labeled Threshold Crossing Events (TCEs). We obtained all the 15,737 TCEs produced by Kepler and utilized in Google's research deep learning method \citep{SV18,Dat19}, where they used a supervised convolutional neural network machine learning architecture that included 2,202 features: 201 features of `local view' and 2,001 features of `global view'. The `global view' represents the entire light curve and the `local view' represents a phase-folded window around the identified transit.

We derived our training set of labeled TCEs from the Autovetter Planet Candidate Catalog for Q1-Q17 DR24 \citep{Cat15,Cou16} hosted at the NASA Exoplanet Archive (\url{https://exoplanetarchive.ipac.caltech.edu/}). We obtained the TCE labels from the catalog's  {\tt av\_training\_set} column, which has three possible values: planet candidate (PC), astrophysical false positive (AFP) and non-transiting phenomenon (NTP). We ignored TCEs with the “unknown” label (UNK). These labels were produced by manual vetting and other diagnostics. We obtained additional data on the TCEs  such as planet number, radius of the planet, interval between consecutive planetary transits, etc., from the MAST TESS archive ( \url{https://archive.stsci.edu/missions-and-data/transiting-exoplanet-survey-satellite-tess}) for data labeling and use in our analysis.

\subsection{Features}
Feature engineering is the process of using data domain knowledge to create features by manipulating the data through mathematical and statistical relations (for examples, see section~\ref{new-features:sec}) of the various components in order to improve the performance of the AI/ML algorithms. The feature engineering process includes  deciding which features to develop, creating the features, checking how the features work with the model, improving the features as needed, and going back to deciding on or creating additional data features until the ML/AI algorithm results are optimized. We applied the feature engineering process on our dataset and created new features in addition to the existing features available in MAST in order to provide more information which will quantify various aspects of the data used by the AI/ML algorithm in the present analysis. We produced a total of 424 features that were used for the analysis.   We chose the combination of features that provided the best results under the capabilities of {\em ThetaRay}'s system, validated in the training step. In the feature engineering process, we tested the effectiveness of different combinations of features under the limits of {\em ThetaRay}'s system.

\subsection{Existing features}
\label{existing_feat:sec}
Additional TCEs Data were downloaded from MAST. We narrowed down the data only to the required fields for the present task, such as the planet number, the radius of the planet, the interval between consecutive planetary transits, etc., and  selected the relevant data from all the fields from ``Data Columns in the Kepler TCE Table” (\url{https://exoplanetarchive.ipac.caltech.edu/docs/API\_tce\_columns.html}) using the visualization of the variables (especially KDE plots, see below). Below is the description of the variables and labels used in our analysis.

\begin{itemize}
\item	{\tt  Unique key} - concatenation of Kepler ID and Planet Number. Kepler ID is a target identification number, as listed in the Kepler Input Catalog (KIC). The KIC was derived from a ground-based imaging survey of the Kepler field conducted prior to launch. The survey's purpose was to identify stars for the Kepler exoplanet survey by magnitude and color. The full catalog of 13 million sources can be searched at the MAST archive. The subset of 4 million targets found upon the Kepler CCDs can be searched via the Kepler Target Search form. \begin{itemize}
 \item Kepler Input Catalog (KIC) \citep{Bro11}.
 \item MAST archive - \url{http://archive.stsci.edu/kepler/kic10/search.php}.
\item Kepler Target Search form - \url{http://archive.stsci.edu/kepler/kepler\_fov/search.php}.
\end{itemize}
\item	{\tt av\_training\_set} - Autovetter Training Set Label. If the TCE was included in the training set, the training label encodes what is believed to be the ``true" classification, and takes a value of either PC, AFP or NTP. The TCEs in the UNKNOWN class sample are marked UNK. Training labels are given a value of NULL for TCEs not included in the training set. For more detail about how the training set is constructed, see Autovetter Planet Candidate Catalog for Q1-Q17 Data Release 24 (KSCI-19091):
\url{https://exoplanetarchive.ipac.caltech.edu/docs/KSCI-19091-001.pdf}.
\item {\tt tce\_prad} - Planetary Radius (Earth radii). The radius of the planet obtained from the product of the planet to stellar radius ratio and the stellar radius.
\item	{\tt tce\_max\_mult\_ev} - Multiple Event Statistic (MES). The maximum calculated value of the MES. TCEs that meet the maximum MES threshold criterion and other criteria listed in the TCE release notes are delivered to the Data Validation (DV) module of the data analysis pipeline for transit characterization and the calculation of statistics required for disposition. A TCE exceeding the maximum MES threshold are removed from the time-series data and the SES and MES statistics recalculated. If a second TCE exceeds the maximum MES threshold then it is also propagated through the DV module and the cycle is iterated until no more events exceed the criteria. Candidate multi-planet systems are thus found this way. Users of the TCE table can exploit the maximum MES statistic to help filter and sort samples of TCEs for the purposes of discerning the event quality, determining the likelihood of planet candidacy, or assessing the risks of observational follow-up. DV module – \url{http://archive.stsci.edu/kepler/manuals/KSCI-19081-001\_Data\_Processing\_Handbook.pdf}
\item {\tt tce\_period} - Orbital Period (days). The interval between consecutive planetary transits.
\item	{\tt tce\_time0bk} - Transit Epoch (BJD) - 2,454,833.0. The time corresponding to the center of the first detected transit in Barycentric Julian Day (BJD) minus a constant offset of 2,454,833.0 days. The offset corresponds to 12:00 on Jan 1, 2009 UTC.
\item {\tt tce\_duration} - Transit Duration (hrs). The duration of the observed transits. Duration is measured from first contact between the planet and star until last contact. Contact times are typically computed from a best-fit model produced by a \citet{MA02} model fit to a multi-quarter Kepler light curve, assuming a linear orbital ephemeris.
\item {\tt tce\_model\_snr} - Transit Signal-to-Noise (SNR). Transit depth normalized by the mean uncertainty in the flux during the transits.
\item {\tt av\_pred\_class} - Autovetter Predicted Classification. Predicted classifications, which are the `optimum MAP classifications.' Values are either PC, AFP, or NTP.
\item {\tt tce\_depth} - Transit Depth (ppm). The fraction of stellar flux lost at the minimum of the planetary transit. Transit depths are typically computed from a best-fit model produced by the \citet{MA02} model fit to a multi-quarter Kepler light curve, assuming a linear orbital ephemeris.
\item {\tt tce\_impact} - Impact Parameter. The sky-projected distance between the center of the stellar disc and the center of the planet disc at conjunction, normalized by the stellar radius.
\item {\tt local\_view} - vector of length 201: a `local view' of the TCE. It shows the shape of the transit in detail (close-up of the transit event).
\end{itemize}

\subsection{Visualization of Kepler Data}

\begin{figure}
\begin{centering}
\includegraphics[width=0.90\textwidth]{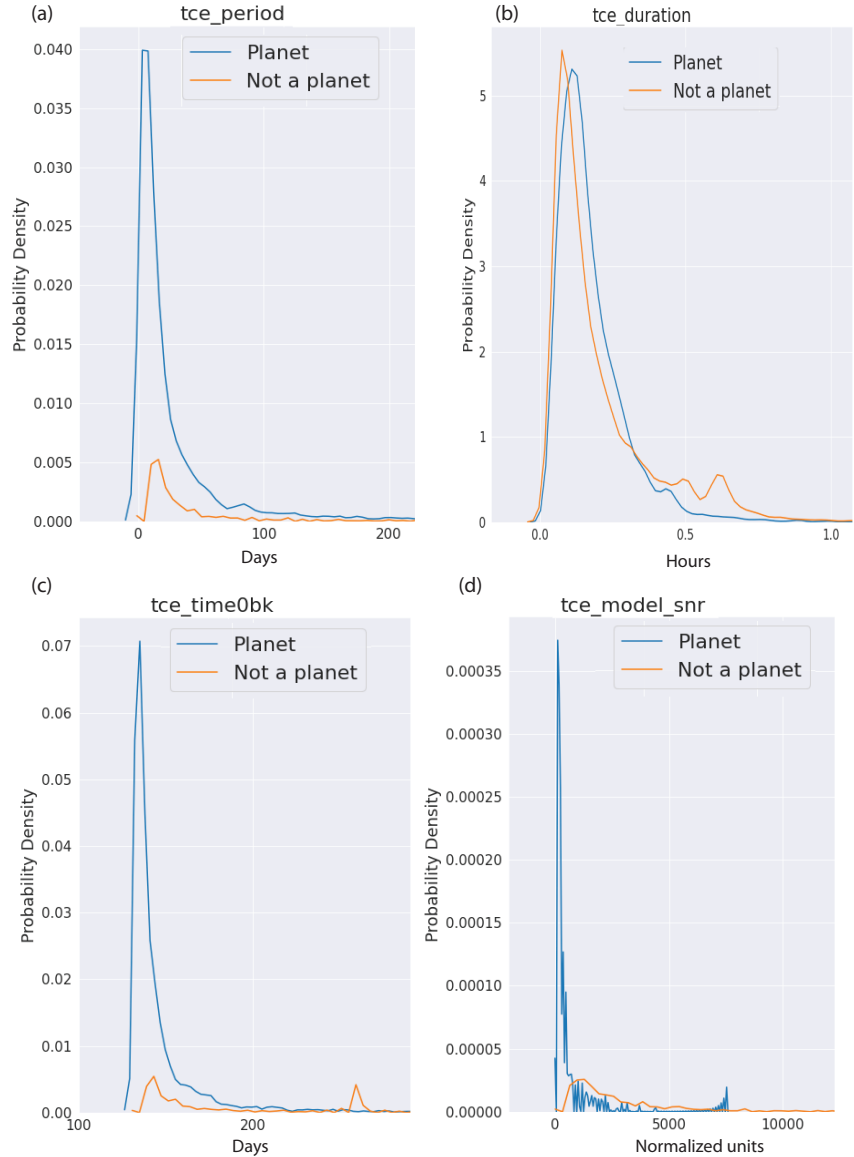}
\vspace{-1cm}\caption{The distributions of the numerical variables using KDE (Kernel Density Estimation) plots where the blue curves are labeled `Planet' and the orange curves are labeled ` Not a planet' from Kepler data. When there is significant difference between the curves, it can be concluded that these features are more significant for planet identification and therefore we have included them in the model. If both curves coincide, it can be concluded that the behavior is not statistically different between the two populations. The plotted variables are (a) {\tt tce\_period}, (b) {\tt tce\_duration}, (c) {\tt tce\_time0bk}, (d) {\tt tce\_model\_snr} (see text for their definitions). }
\label{dist:fig}
\end{centering}
\end{figure}

We investigated the Kepler data and visualized the variables with Pandas package in Python. For example, we visualize the distributions of the numerical variables per class using KDE (Kernel Density Estimation) plots. In Figure~\ref{dist:fig} we show several interesting examples with a gap between the curves labeled `Planets' and `Not planets'  as identified by {\em ThetaRay} system and validated by the Kepler data training set. It can be concluded that these features are significant in candidate exoplanet identification and therefore we have included them in the model. If both curves coincide, it can be concluded that the behavior is the same for label `planets' and `not planets', and so we chose not to include these features in the model.

\begin{figure}
\begin{centering}
\includegraphics[width=6in]{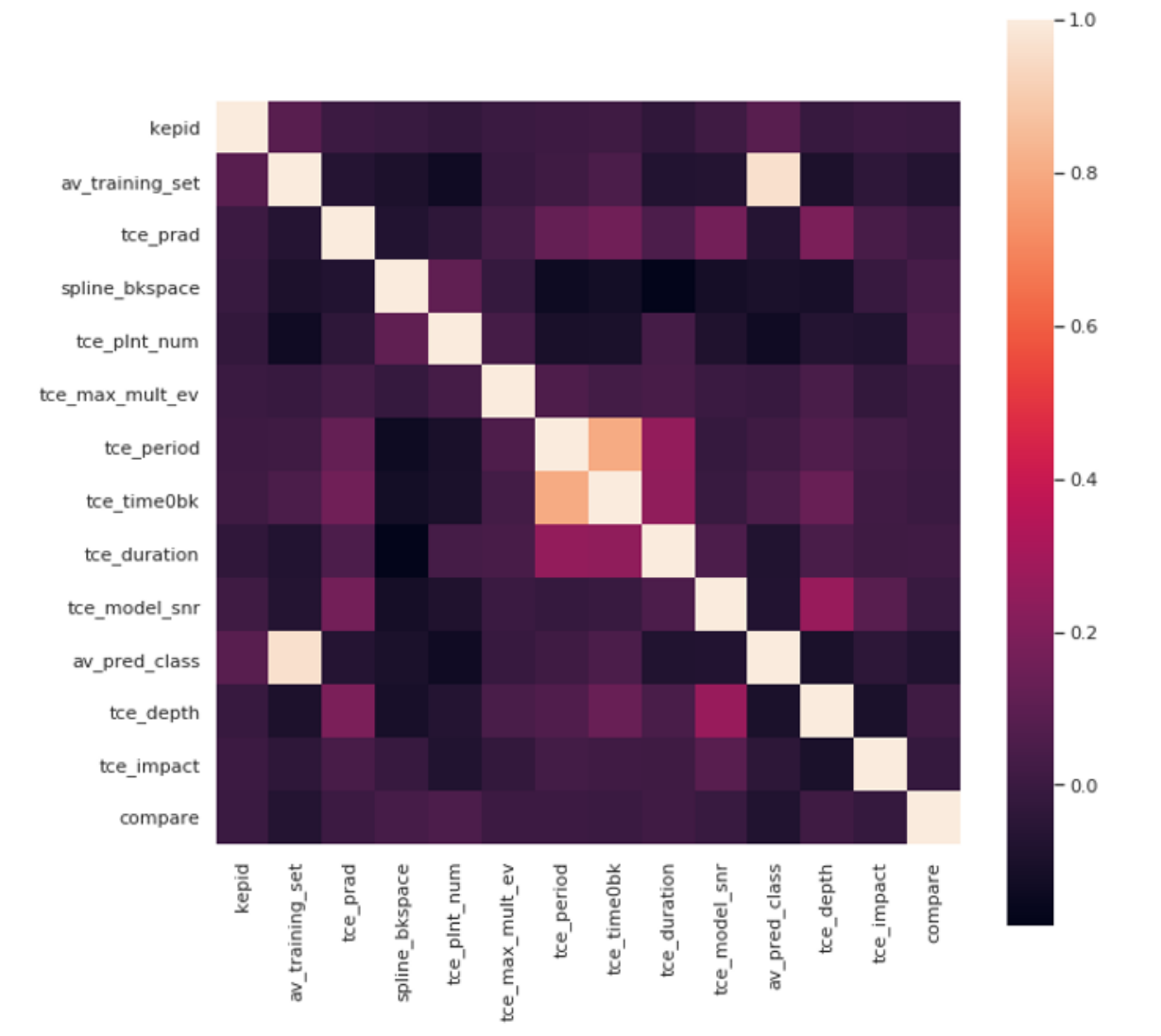}
\caption{The `Heat map' of some of the features (or parameters) used in the {\em ThetaRay} algorithm. The intensity scale indicates the magnitude of the correlation between the features that facilitates determining the dimensionality of the dataset (see text).}
\label{heat:fig}
\end{centering}
\end{figure}

Another example of our analysis is demonstrated in the `heat map', which is basically a color-coded matrix, where a correlation value between the variable of features is used to color each cell of the matrix to represent the relative value of that cell.  If there is a high correlation between any variables, the dimension of the data can be reduced. The various features are labeled on the axes. Obviously, the features on the main diagonal that indicate identity correlation are light colored. It is evident from the `heat map' shown in  Figure~\ref{heat:fig} that most off-diagonal features are weakly correlated. The only significant off-diagonal correlations are between
{\tt av\_training\_set } - the training labels, i.e., if the TCE was included in the training set, the training label encodes what is believed to be the ``true" classification, and {\tt av\_pred\_class} - predicted classifications, which are the optimum MAP (maximum a posteriori) classifications. In fact, this field does not provide analysis information for the data but is used as forensic feature. The forensic features are not included directly in the analysis but  provide supplementary information about the data useful for the investigation of the  analysis. Some artificial correlation is also evident between the {\tt tce\_time0bk} - transit epoch (BJD), and {\tt tce\_period} - Orbital Period (days).

\subsection{New Features}
\label{new-features:sec}
New features  were developed based on the original data set from Kepler that was obtained from MAST to optimize the analysis with {\em ThetaRay} algorithm. These features were constructed from the original dataset as described below using the phase-folded ``Local View'' light curves \citep[see, e.g.,][]{SV18}.
\begin{itemize}
\item	{\tt global\_view} - the original vector of length 2001 or a `global view' of the TCE that shows the characteristics of the light curve over an entire orbital period. Because of the size limitations of the {\em ThetaRay}’s system, we performed dimension reduction. We represented groups of 20 columns in the `global view' by computing the average and standard deviation of those columns. We have a total of 200 new “global\_view” features.
\item	{\tt spline\_bkspace} - the breakpoint spacing in time units, used for the best-fit spline. We chose the optimal spacing of spline breakpoints for each light curve by fitting splines with different breakpoint spacings, calculating the Bayesian Information Criterion (BIC, \citet{Sch78}) for each spline, and choosing the breakpoint spacing that minimized the BIC. Below, is a brief description of the new features that were computed for each TCE ``Global View'' and ``Local View'' light curves:
\item	{\tt loc\_mean} – average of the ``Local View” light curve.
\item	{\tt loc\_std} - standard deviation of the ``Local View” light curve.
\item	{\tt loc\_25\%} -25\% percentile of the ``Local View” light curve.
\item	{\tt loc\_75\%} - 75\% percentile of the ``Local View” light curve.
\item	{\tt loc\_max} – max value of the ``Local View” light curve.
\item	{\tt glob\_mean} – average of the original ``Global View” light curve.
\item	{\tt glob\_std}- standard deviation of the original ``Global View” light curve.
\item	{\tt glob\_25\%} - lower percentage of the original ``Global View” light curve.
\item	{\tt glob\_75\%} - upper percentage of the original ``Global View” light curve.
\item	{\tt glob\_max} – max value of the original ``Global View” light curve.
\item	{\tt zScore\_loc\_min} – minimum value of the Z-Score on the ``Local View” light curve with window of 10.
\item	{\tt zScore\_loc\_max} – maximum value of the Z-Score on the ``Local View” light curve with window of 10.
\item	{\tt zScore\_glob\_min} – minimum value of the Z-Score on the ``Global View” light curve with window of 100.
\item	{\tt zScore\_glob\_max} – maximum Z of the-Score on the ``Global View” light curve with window of 100.
\end{itemize}

\section{TESS Satellite Data Analysis}
\label{TESS:sec}
\subsection{Preprocessing the Data}
We obtained 10,803 light curves of TCEs produced by the TESS mission from MAST ( \url{http://archive.stsci.edu/}). We wanted to use the same model we built based on Kepler's data, in order to find potential exoplanets (anomalies) in the new data from TESS. For using the same models for the two different satellites, we must convert the TESS data to the same structure as Kepler data. Therefore, we performed additional steps to prepare the light curves to be used as inputs to our system. We generated a set of TFRecord files for the TCEs. Each file contains {\tt global\_view}, {\tt local\_view} and {\tt spline\_bkspace} representations like in Kepler. We also created in python the following data  files:
\begin{itemize}
\item	{\tt global\_view} - Vector of length 2001 that shows the characteristics of the light curve over an entire orbital period.
\item	{\tt local\_view} - Vector of length 201 that shows the shape of the transit in detail (phase-folded close-up of the transit event).
\item	{\tt more\_features} - includes
\begin{itemize}
\item {\tt ticid} - TESS ID of the target star.
\item {\tt planetNumber} - TCE number within the target star.
\item {\tt planetRadiusEarthRadii} - has the same meaning as the field of {\tt tce\_prad} in Kepler  data.	
\item {\tt spline\_bkspace}, the breakpoint spacing in time units, used for the best-fit spline - same as in Kepler  data.
\item {\tt mes} - same meaning as {\tt tce\_max\_mult\_ev}  in Kepler data.
\item {\tt orbitalPeriodDays} - same meaning as  {\tt tce\_period} in Kepler data.
\item  {\tt transitEpochBtjd} - same meaning as {\tt tce\_time0bk} in Kepler data.
\item {\tt transitDurationHours} - same meaning as  {\tt tce\_duration} in Kepler data.
\item  {\tt transitDepthPpm} - same meaning as {\tt tce\_depth} in Kepler Data.
\item  {\tt minImpactParameter} - same meaning as {\tt tce\_impact} in Kepler data.
\end{itemize}
TESS data is unlabeled, so {\tt av\_training\_set} and {\tt av\_pred\_class} fields do not exist in the TESS data, therefore, we filled these fields with zeros. {\tt tce\_model\_snr} feature was calculated value by the ratio of {\tt transitDepthPpm} and {\tt transitDepthPpm\_err}.
\item	{\tt Describe files} - includes count, mean, std, min, max, 25\% percentile, median (50\%), 75\% percentile. These quantities were computed on each original data row from the {\tt global\_view} and {\tt local\_view} files and on each scaling row of these files.
\end{itemize}

Following the generation of the dataset in the form of Comma Separated Values (CSVs), we applied the same manipulation on {\tt global\_view}, as in Kepler data, in order to reduce the dimensions, and used the analogous 424 features produced from TESS data as in Kepler data, for the analysis on {\em ThetaRay}'s system. Following this step, we applied the  {\tt Detection} algorithm on TESS data according to the saved model from Kepler and used the results for classification and mapping of TESS light curve TCEs data.

\subsection{Results: Transiting Exoplanet Detection}
\label{Results:sec}
The first results of the {\em ThetaRay} algorithm that was applied to TESS data TCEs in December 2019 produced around 90 preliminary { targets} that were further manually vetted { removing obvious non-planets and 12 cases due to momentum dumps of the TESS satellite \citep[see, e.g.,][]{Van19}, and thus} reducing the number of { targets for analysis} by about  a factor of two. Local view light-curves were used together with planetary candidate parameters to vet the algorithm's output. In the manual vetting the physical parameters, such as non-typical `local view' light curves (i.e., v-shapes, and other non-planetary periodic features), extremely large planetary radius ($r\gtrsim 20R_{Earth}$), and very low signal-to-noise were used. The parameters for the remaining 61 identified EPCs by the {\em ThetaRay} system form the TESS database of 10,803 TCE's are given in Table~\ref{planets:tab}, and the disposition of each identified EPC is provided in Table~\ref{disposition:tab}. It is evident that { three} identified cases are EPCs, while the remaining cases are known exoplanets, planetary candidates, and signals due to other effects, { such as} off-target light-curves, and scattered light.  Cross-checking our results  with the TESS planet candidate list {\tt https://tev.mit.edu/data/collection/193/}, matches 19 cases in our table  of planet candidates and confirmed planes as indicated in Table~\ref{disposition:tab}.  While these 19 cases are not new EPCs, the overlap with the MIT planet candidate list improves confidence in the performance of the {\em ThetaRay} algorithm in identifying EPCs.

\begin{figure}
\begin{centering}
\vspace{-4cm}\includegraphics[width=\linewidth]{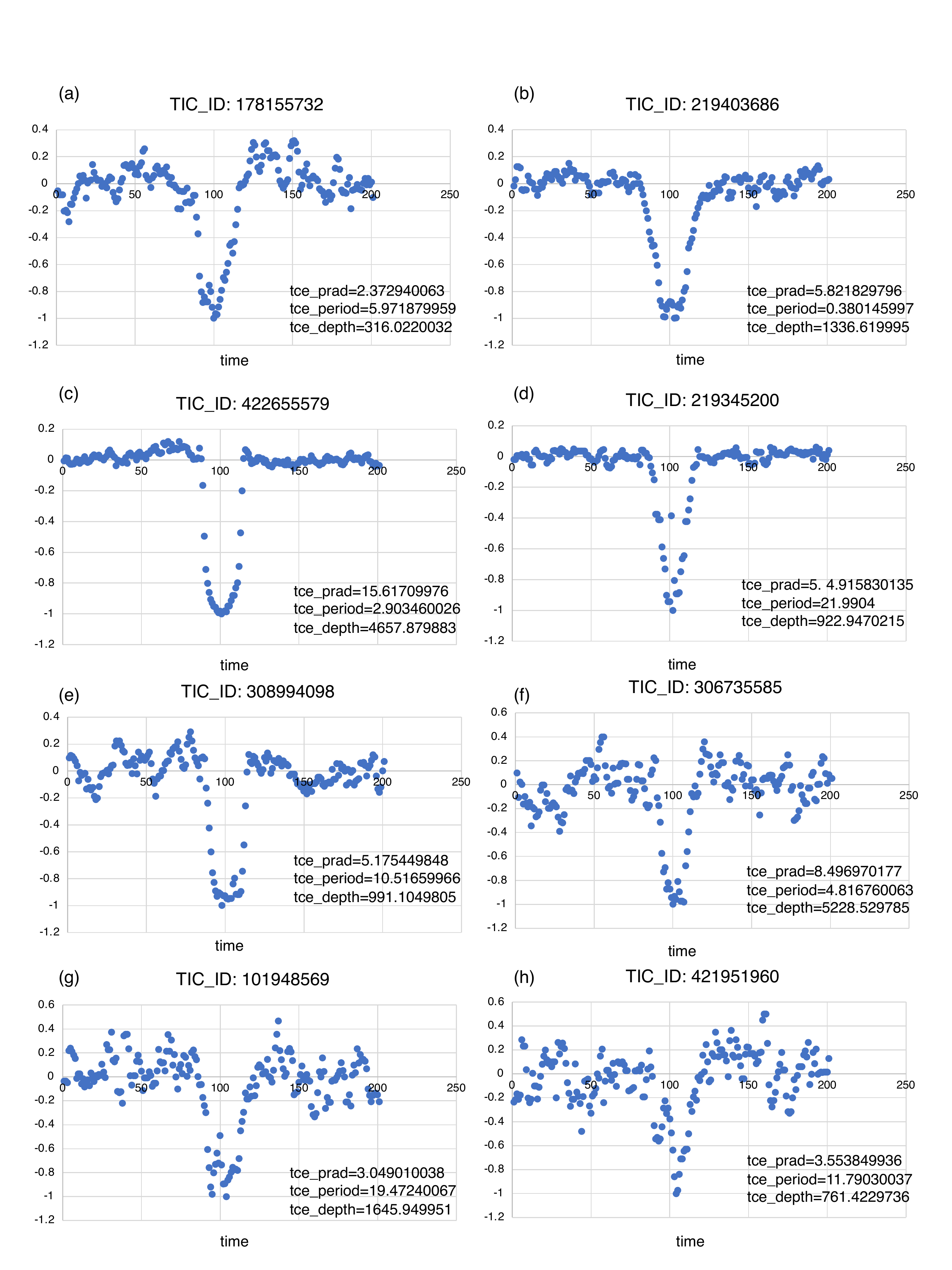}
\vspace{-1.5cm}\caption{``Local view'' normalized phase-folded light-curves of six objects selected from Table~\ref{planets:tab} representing different types of detections by the {\rm ThetaRay} algorithm. The parameters {\tt tce\_prad} the radius in terms of $R_{Earth}$, {\tt tce\_period} in days, {\tt tce\_depth} in ppm, indicated on the corresponding panels. The typical eclipsing exoplanetary light curve temporal shape structure is evident. (a) TIC\_ID: 178155732: known planet HR 858 c, (b) TIC\_ID: 219403686: possible new planet candidate, (c) TIC\_ID: 422655579: known planet WASP-71 b, (d) { TIC\_ID: 219345200:} possible new planet candidate,  (e)  TIC\_ID: 308994098 is a known planet candidate TOI 790, (f) TIC\_ID: 308994098: known planet candidate TOI 790, (g) { TIC\_ID: 306735585: possible new planet candidate (but see comment in Table~\ref{disposition:tab})}, (h) { TIC\_ID: 421951960:} possible new planet candidate.}
\label{local:fig}
\end{centering}
\end{figure}

In Figure~\ref{local:fig} we show the Local View light curves of eight selected light curves for the objects identified using the {\em ThetaRay} algorithm as potential EPCs. The TESS input catalog ID number (TIC\_ID), along with several parameters ({\tt tce\_prad, tce\_period, tce\_depth} defined in section~\ref{existing_feat:sec}) for the identified EPC  are indicated on each panel. The eight panels in Figure~\ref{local:fig} show typical eclipsing exoplanetary light curve temporal shape structure. In particular TIC\_ID: 178155732 was identified by the algorithm, and is a known planet HR 858 c, TIC\_ID: 422655579 is a known planet WASP-71 b, and TIC\_ID: 308994098 is a known planet candidate TOI 790. While these objects are not new EPCs, their identification provides  validation of the ML algorithm. The objects TIC\_ID: { 219345200}, 219403686, { 306735585, 421951960}, are shown in Figure~\ref{local:fig} are possible new EPCs identified by the algorithm { (with some reservation with respect to TIC\_ID: 306735585 indicated in Table~\ref{disposition:tab})}.

We note that Table~\ref{planets:tab} only three case with planetary radius ({\tt tce\_prad}) or $r_p<2R_{Earth}$  (TIC\_ID 307210830 known planet  L98-59 b, TIC\_ID 259377017 known planet TOI 270 b, and TIC\_ID 305751503 new EPC), and a total of 17 EPCs identified with $r_p<3R_{Earth}$. Another 13 identified EPCs were similar in size or larger than Jupiter with $r_p\ge 10R_{Earth}$, among them 6 were previously identified planets or TOIs, and 3  due instrumental effects, and 3 are new EPCs. 
We find the following properties of the data in  Table~\ref{planets:tab} \begin{itemize}
\item The orbital periods ({\tt tce\_period}) of the identified EPCs range from 0.25d to just under 22d.
\item The transit depth ({\tt tce\_depth}) varied by about an order of magnitude in the range $305-63679$ ppm with the signal-to-noise in the range $\sim5-92$ ppm.
\item  The impact parameter was in the range $\sim0.01-0.99$.
\item The duration of the transits ({\tt tce\_duration}) was in the range $0.016-0.4461$ d.
\item In 8 cases the identified EPCs suggest multiple planetary systems with 2 and 3 planets, with two new EPCs.

\end{itemize}

\begin{table}
\tiny
\vspace{-3cm}\caption{\small Some of the parameters (see text) of exoplanetary candidates identified by the {\em ThetaRay} system in TESS mission data archive { (see text). Column `s' indicates the sector(s).}}
\hspace{-1cm}
\begin{tabular}{lclllllllll}
\hline
TIC\_ID	& \# p  & s & tce\_prad & tce\_max&  tce\_period & tce\_time0bk & tce\_duration & tce\_model & tce\_depth	&   tce\_impact  \\
          	&          &    &                & \_mult\_ev  &                &                      &                        &                   & 	               &      \\
\hline
101948569 & 1 & 2 & 3.049010038 & 11.03339958 & 19.47240067 & 1360.109985 & 3.47104 & 11.74638202 & 1645.949951 & 0.303943992 \\
102195674 & 1 & 9 & 20.62459946 & 68.38349915 & 4.378769875 & 1547.459961 & 3.8184 & 67.04560811 & 30425.69922 & 0.00999983 \\
117979694 & 1 & 5 & 2.017019987 & 8.018360138 & 0.991541028 & 1438.709961 & 1.99801 & 10.88359612 & 662.7739868 & 0.557057977 \\
120916706 & 1 & 3 & 5.22453022 & 10.40649986 & 0.556737006 & 1386.170044 & 0.879121 & 11.14050062 & 63679.10156 & 0.00999983 \\
124573851 & 1 & 9 & 3.842240095 & 9.629090309 & 3.948679924 & 1546.140015 & 3.42051 & 11.05563987 & 754.0189819 & 0.935890973 \\
164767175 & 2 & 3 & 2.407740116 & 9.309049606 & 10.76780033 & 1393.079956 &  2.81587 & 8.523747457 & 650.6099854 & 0.202473998 \\
167418903 & 1 & 11 & 10.36499977 & 20.16550064 & 21.96240044 & 1599.300049 & 1.75464 & 17.57117062 & 8278.05957 & 0.844699979 \\
167603396 & 1 & 1-3 & 2.751130104 & 8.034460068 & 14.37919998 & 1365.140015 & 6.48194 & 8.00441099 & 1444.5 & 0.269908011 \\
170849515 & 1 & 5 & 10.83209991 & 16.08620071 & 1.941280007 & 1438.150024 & 1.80662 & 20.93755124 & 37131.69922 & 0.00999983 \\
172464366 & 1 & 6 & 19.09110069 & 79.50800323 & 2.921689987 & 1470.050049 & 3.17791 & 77.20788698 & 17562.40039 & 0.552250981 \\
178155732 & 1 & 1-36 & 2.372940063 & 10.47840023 & 5.971879959 & 1415.630005 & 3.12355 & 12.1434367 & 316.0220032 & 0.226411998 \\
178819686 & 2 & 10 & 3.705120087 & 9.265979767 & 12.27630043 & 1572.969971 & 3.65638 & 10.82845891 & 976.2700195 & 0.603969991 \\
200591694 & 1 & 6 & 4.385819912 & 8.523739815 & 13.58699989 & 1470.150024 & 2.49368 & 9.010778093 & 4702.040039 & 0.00999983 \\
219345200 & 1 & 4 & 4.915830135 & 44.37039948 & 21.9904 & 1414.579956 & 4.44366 & 15.23623284 & 922.9470215 & 0.342040002 \\
219379012 & 1 & 6 & 4.340690136 & 15.25220013 & 1.546159983 & 1469.709961 & 1.34256 & 17.6102274 & 1064.26001 & 0.692296982 \\
219403686 & 1 & 6 & 5.821829796 & 22.26140022 & 0.380145997 & 1468.579956 & 0.692123 & 29.63871326 & 1336.619995 & 0.679122984 \\
254700590 & 1 & 12 & 2.881239891 & 7.790110111 & 11.77639961 & 1625.609985 & 4.64722 & 7.599164669 & 543.3380127 & 0.651623011 \\
259377017 & 3 & 5 & 1.372750044 & 9.26651001 & 3.359859943 & 1387.089966 & 1.40678 & 8.651269232 & 1034.959961 & 0.463200003 \\
260647166 & 2 & 11 & 2.615370035 & 11.52190018 & 15.85879993 & 1606.160034 & 3.89359 & 11.53021367 & 829.5319824 & 0.00999983 \\
270677759 & 1 & 11 & 9.437470436 & 14.45300007 & 9.129110336 & 1597.199951 & 4.72452 & 14.20179783 & 8185.049805 & 0.805234015 \\
293844307 & 1 & 8 & 4.286910057 & 10.61699963 & 14.76070023 & 1525.550049 & 3.56351 & 10.77202809 & 1557.839966 & 0.201021999 \\
296972860 & 1 & 9 & 11.50829983 & 17.99360085 & 12.45800018 & 1549.77002 & 3.38762 & 12.5118297 & 2250.340088 & 0.990104973 \\
300560295 & 2 & 12 & 5.527339935 & 7.324260235 & 6.884880066 & 1625.630005 & 2.36212 & 7.720771781 & 1999.51001 & 0.754741013 \\
302971087 & 1 & 6 & 1.856520057 & 7.757989883 & 17.5326004 & 1469.609985 & 1.81434 & 6.415764637 & 1347.23999 & 0.306959003 \\
303051566 & 1 & 1 & 3.854789972 & 8.216239929 & 19.6228 & 1328.569946 & 1.49412 & 6.706231497 & 569.2410278 & 0.0374793 \\
306735585 & 1 & 1-6 & 8.496970177 & 12.65380001 & 4.816760063 & 1414.079956 & 2.29767 & 11.34648808 & 5228.529785 & 0.893122017 \\
307210830 & 3 & 11 & 0.869957983 & 11.3927002 & 2.253309965 & 1598.23999 &1.01187 & 10.71902329 & 723.9060059 & 0.501681983 \\
307467401 & 1 & 6 & 4.158410072 & 44.08229828 & 9.587329865 & 1475.709961 & 6.74873 & 25.5473955 & 1375.969971 & 0.83335799 \\
308994098 & 1 & 9 & 5.175449848 & 19.70980072 & 10.51659966 & 1552.050049 & 10.7063 & 18.35474766 & 991.1049805 & 0.707704008 \\
309619055 & 1 & 9 & 10.75909996 & 20.72480011 & 10.55350018 & 1604.969971 & 4.59559 & 11.93659993 & 9440.740234 & 0.894083023 \\
311925584 & 1 & 11 & 14.58570004 & 19.55500031 & 3.645989895 & 1599.189941 & 1.35795 & 13.91252537 & 8853.110352 & 0.96806401 \\
313676015 & 1 & 11 & 8.788999557 & 12.18910027 & 12.95289993 & 1604.140015 & 5.25315 & 11.46735907 & 3192.100098 & 0.987146974 \\
322900369 & 1 & 1-36 & 7.504670143 & 95.60189819 & 3.126100063 & 1493.890015 & 12.7532 & 92.18038804 & 5151.629883 & 0.570958972 \\
335452175 & 1 & 11 & 15.31970024 & 58.07910156 & 15.49790001 & 1601.26001 & 3.22331 & 60.38542514 & 8295.75 & 0.990104973 \\
355509914 & 1 & 1-36 & 10.77110004 & 7.868070126 & 1.738260031 & 1326.119995 & 1.29881 & 8.795540018 & 18978.40039 & 0.200622007 \\
365638739 & 1 & 6 & 3.920919895 & 8.19054985 & 17.21910095 & 1470.390015 & 3.62824 & 7.832664711 & 2136.790039 & 0.301687002 \\
370228465 & 2 & 1-2 & 4.152969837 & 7.710509777 & 12.32479954 & 1357.910034 & 15.6542 & 7.614795753 & 4594.089844 & 0.0166821 \\
392113518 & 1 & 11 & 5.140649796 & 11.11190033 & 4.366240025 & 1569.849976 & 5.57279 & 11.75887721 & 847.0930176 & 0.416761011 \\
393474846 & 1 & 10 & 1.801599979 & 7.123549938 & 11.05160046 & 1578.839966 & 2.60486 & 6.741978404 & 305.5039978 & 0.461097002 \\
396902326 & 1 & 9 & 6.775030136 & 13.27799988 & 15.89120007 & 1545.609985 & 7.86604 & 14.09583373 & 4045.179932 & 0.987653017 \\
406872253 & 1 & 12 & 2.106899977 & 7.377220154 & 0.246232003 & 1625.170044 & 0.392131 & 6.283381761 & 394.3840027 & 0.165623993 \\
410214984 & 3 & 1-36 & 4.706439972 & 12.51220036 & 8.135899544 & 1332.349976 & 3.28862 & 6.290396653 & 4277.52002 & 0.144591004 \\
421951960 & 1 & 1-36 & 3.553849936 & 8.082489967 & 11.79030037 & 1332.48999 & 1.60633 & 8.191518961 & 761.4229736 & 0.390895009 \\
422655579 & 1 & 4 & 15.61709976 & 27.75729942 & 2.903460026 & 1413.140015 & 5.02954 & 45.99784581 & 4657.879883 & 0.00999983 \\
423275733 & 1 & 8 & 17.97360039 & 24.79450035 & 2.052979946 & 1518.689941 & 2.65216 & 36.4699389 & 10176.90039 & 0.745383978 \\
447061717 & 1 & 10 & 2.608789921 & 21.75650024 & 9.204919815 & 1569.719971 & 3.24093 & 18.51823286 & 4082.899902 & 0.0298279 \\
453767182 & 1 & 12 & 2.725820065 & 7.798190117 & 10.76249981 & 1626.119995 & 2.93366 & 7.410331196 & 13611 & 0.027590601 \\
455278250 & 1 & 8 & 7.306509972 & 36.73529816 & 15.60929966 & 1521.51001 & 5.72957 & 28.64738963 & 2824.179932 & 0.82368201\\
\hline
\end{tabular}

\label{planets:tab}
\end{table}

\begin{table}
\tiny
\vspace{-3cm}\caption{\small The disposition of exoplanetary candidates identified by the {\em ThetaRay} system in TESS mission data archive.}
\hspace{-1cm}
\begin{tabular}{ll}
\hline
TIC\_ID	& Disposition     \\
\hline
101948569 & known planet candidate TOI 240 \\
102195674 & known planet candidate TOI 668 \\
117979694 & known planet candidate TOI 424/off target \\
120916706 & known planet candidate TOI 263 \\
124573851 & known planet candidate TOI 669 \\
164767175 & known planet candidate TOI 266 \\
167418903 & known planet candidate (detected at the wrong 2x period) TOI 746 \\
167603396 & transit/possibly hartbeat star\footnotemark \\
170849515 & known planet candidate TOI 555 \\
172464366 & known planet candidate (confirmed to be an eclipsing binary) TOI 478 \\
178155732 & known planet HR 858 c \\
178819686 & known planet candidate TOI 763 \\
200591694 & off-target \\
219345200 &  { possible new planet candidate} \\
219379012 & off-target \\
219403686 & possible new planet candidate  \\
254700590 & possibly single transit? \\
259377017 & known planet TOI 270 \\
260647166 & known planet candidate TOI 1233 \\
270677759 & off-target \\
293844307 & { off-target} \\
296972860 &  {signal apparently due to an asteroid in the solar system passing} \\
                  &   { near the target star affecting the background light subtraction\footnotemark}\\
300560295 & { off-target} \\
302971087 &  { atypical transit shape}\\
303051566 & possibly single transit? \\
306735585 & possible planet but ghost diagnostic is high$^3$ \\
307210830 & known planet L 98-59  \\
307467401 & scattered light event$^4$  \\
308994098 & known planet candidate TOI 790 \\
309619055 & off-target \\
311925584 &  { significant odd/even transit depth difference}\\
313676015 &  { single transit/off-target} \\
322900369 & clear eclipsing binary with tidal distortion right after eclipse \\
335452175 & off-target \\
355509914 & known planet candidate TOI 202/Possibly off-target\footnotemark \\
365638739 &  single transit/off-target \\
370228465 & possibly single transit? \\
392113518 & secondary eclipse/off-target\\
393474846 & { stellar variability/not transit-shaped light curve} \\
396902326 &  { off-target}\\
406872253 &  { off-target} \\
410214984 & signal due to known planet DS Tuc Ab (on nearby star)  \\
421951960 &  { possible new planet candidate} \\
422655579 & known planet WASP-71 b \\
423275733 & known planet WASP 142 b \\
447061717 & known planet candidate TOI 1231 \\
453767182 & off target \\
455278250 & scattered light \\
\hline
\end{tabular}\\
{ $^1$see, e.g., \citet{Shp16}.}  \\
{ $^2$see, {\tt http://vo.imcce.fr/webservices/skybot/?forms=conesearch}.}  \\
{ $^3$see, \citet{Bry13} and \citet{Jen16} for the analysis method description.}  \\
{ $^4$see, {\tt https://docs.lightkurve.org/tutorials/2-creating-light-curves/2-3-removing-scattered-light-using-regressioncorrector.html} \citet{Dal20,hattori21}.}

\label{disposition:tab}
\end{table}

\section{Discussion and Conclusions} \label{DC:sec}

The TESS satellite provides observations of a large number (200,000) of stellar light curves with high photometric precision over the whole sky, divided in observing sectors, with the aim of detecting transiting Earth-sized planets. The stellar objects were selected to represent the brightest and closest to our solar system. The large dataset of nearly 27 gigabytes per day is then processed in the science data pipeline providing nearly 11,000 TCE's as of the time of writing this paper. Further analysis of the TCEs is required to find confirmed examples of exoplanets, or exoplanetary candidates for more in-depth processing. However, evidently this  formidable data analysis task is difficult, if not impossible to carry out manually. A feasible approach for the TESS data analysis is based on automated identification  techniques that were developed recently, customized for transiting exoplanetary candidates identification, EPCs, utilizing AI/ML methods based on DL neural networks machine learning methods combined with anomaly identification methods reported the present study. These EPCs could be than vetted further with targeted observations and data analysis.

In this study we apply a novel algorithm developed by {\em ThetaRay, Inc.} for cybersecurity and anomaly identification in financial systems to help detected exoplanets in TESS data. This is the first attempt to do this task in an unsupervised or semi-supervised machine learning. The advantage of this AI/ML system over other unsupervised ML methods is the combination of several algorithms, is described in this paper (Section~\ref{ML:sec}), and the direct application to any large dataset that contain possibly small number of target data points (`anomalies'). We apply the system to TESS observations of TCE's in search of transiting exoplanet signatures in the large TCE dataset. For the training set of the ML algorithm, we used the Kepler exoplanet TCE's validated with confirmed exoplanet dataset. By applying the trained {\em ThetaRay} algorithm to TESS TCE's we { identified 48 targets for further analysis}  in wide range of sizes from below Earth's radius to super-Jupiter's radii, and planetary periods ranging from 0.38d to $\sim$20d. { Further manual vetting resulted in the identification of three new exoplanetary candidates (see Table~\ref{disposition:tab}).} We demonstrate that the combination of DL neural networks with anomaly identification mathematical techniques  provide an efficient AI/ML algorithm for the rapid automated search of transiting exoplanet candidates light curves. Although, we find that we need to apply manual vetting to reduce the number of obvious false-positives, the total number of EPCs identifications is manageable for secondary manual vetting of the relatively small number of light-curves, and this approach provides the desired classification of the results. In future applications, the {\em ThetaRay} algorithm could be further optimized for transiting exoplanets light-curves classification helped the identification of EPCs, by including, for example, informed ML steps, potentially reducing further the false-positive rate in this application and providing a new tool for analyzing TESS data.


\section*{Acknowledgment} We acknowledge the use of public TOI Release data from pipelines at the TESS Science Office and at the TESS Science Processing Operations Center. This paper includes data collected by the TESS mission, which are publicly available from the Mikulski Archive for Space Telescopes (MAST). The resources for this research were provided by {\em ThetaRay, Inc.} LO would like to acknowledge the hospitality of the Department of Geosciences, Tel Aviv University. { We thank the referee for the invaluable input that helped improve this paper, and with assistance in vetting the TCE targets.}

\bibliographystyle{elsarticle-harv} 
\bibliography{ofman_planets}

\end{document}